\documentclass[12pt]{iopart}

\begin{document}

\title{Non-oscillating solutions to uncoupled Ermakov systems and the semiclassical
limit}
\author{A.\ Matzkin}

\begin{abstract}
The amplitude-phase formulation of the Schr\"{o}dinger equation is
investigated within the context of uncoupled Ermakov systems, whereby the
amplitude function is given by the auxiliary nonlinear equation. The
classical limit of the amplitude and phase functions is analyzed by setting
up a semiclassical Ermakov system.\ In this limit, it is shown that
classical quantities, such as the classical probability amplitude and the
reduced action, are obtained only when the semiclassical amplitude and the
accumulated phase are non-oscillating functions respectively of the space
and energy variables. Conversely, among the infinitely many arbitrary exact
quantum amplitude and phase functions corresponding to a given wavefunction,
only the non-oscillating ones yield classical quantities in the limit $\hbar
\rightarrow 0$.
\end{abstract}

\address{Department of Physics and Astronomy, University College London,
Gower Street, London WC1E 6BT, UK}


\maketitle

\section{Introduction}

Systems of the form 
\begin{eqnarray}
\partial _{t}^{2}u(t)+k^{2}(t)u(t) &=&\frac{1}{\rho u^{2}(t)}Y(\alpha
(t)/u(t))  \label{1} \\
\partial _{t}^{2}\alpha (t)+k^{2}(t)\alpha (t) &=&\frac{1}{u\alpha ^{2}(t)}%
Z(u(t)/\alpha (t)),  \label{2}
\end{eqnarray}
where $Y$ and $Z$ are arbitrary functions of their arguments, are
generically known as Ermakov systems. They are characterized by the
existence of a first integral, the Ermakov (or Lewis-Ray-Reid) invariant
linking the solutions of Eqs. (\ref{1}) and (\ref{2}), thereby giving rise
to the so-called nonlinear superposition principle. Intensive studies of
their properties, such as their linearization \cite{haas and goedert99} or
their generalization to higher dimensions \cite{schief et al96,kaushal
etal97} have been undertaken to extend the remarkable results concerning
uncoupled systems, ie $Y(\zeta )=0$ and $Z(\zeta )=a^{2}\zeta $ where $a$ is
a constant, obtained by Ray and Reid \cite{ray and reid79,ray reid80}.

Although the paradigm in physical applications of uncoupled Ermakov systems
has been the classical linear time-dependent harmonic oscillator ($t$ being
the time variable), where the full power of Hamiltonian structure \cite
{lewis etal92}, Lagrangian mechanics and Noether symmetries \cite{simic00}
have been employed, it has also been remarked that uncoupled systems link
the time-independent linear Schr\"{o}dinger equation to a nonlinear
'auxiliary' equation in the following way: 
\begin{eqnarray}
\hbar ^{2}\partial _{x}^{2}u(x)+p^{2}(x)u(x) &=&0  \label{3a} \\
\hbar ^{2}\partial _{x}^{2}\alpha (x)+p^{2}(x)\alpha (x) &=&\frac{\hbar
^{2}a^{2}}{\alpha ^{3}(x)},  \label{3b}
\end{eqnarray}
where $x$ refers to the space variable, and $p(x,E)$ to its conjugate
momentum in classical mechanics. $E$ is the energy, assumed to be conserved,
of the system. We have recently shown \cite{matz2000} that the non-linear
equation (\ref{3b}) corresponds to the equation for the amplitude function $%
\alpha (x)$ in the amplitude-phase formulation of the Schr\"{o}dinger
equation, which arises by performing a so-called Milne transform on the
wave-function; the phase function $\phi (x)$ is obtained by integrating the
relation $\partial _{x}\phi =\alpha ^{-2}$ (see Sec. 2).

We shall be concerned throughout this paper by the oscillatory properties of
the solutions of the uncoupled Ermakov system formed by Eqs. (\ref{3a}) and (%
\ref{3b}) in the specific case of a potential energy function having a
single minimum. Actually, since Eq. (\ref{3a}) defines a Sturm-Liouville
problem, the oscillatory properties of the solutions are well-known \cite
{levitan sargsjan91}, and we will restrict our analysis to the oscillations
as a function of $x$ and $E,$ of the amplitude and phase functions. Our
first aim will be to show that, absolutely smooth, that is non-oscillating
amplitude-phase functions can be constructed. Let us recall that the
amplitude-phase formulation of the one-dimensional Schr\"{o}dinger equation
is frequently used in quantum scattering theories that explicitly include
closed channels, such as quantum defect theory (see \cite{matz2000} and
references therein)$.$ In those situations, it is of prime importance for $%
\alpha (x,E)$ to be a smooth function of both the space (usually radial)
coordinate $x$ and the energy $E, $ since the scattering parameters (for
example the phase-shifts) are defined in terms of amplitude-phase functions.
However, by the principle of nonlinear superposition (Sec. 2.1), $\alpha $
may be expressed in terms of 2 independent solutions $u_{1}$ and $u_{2}$ of
Eq. (\ref{3a}) -- solutions which as known oscillate between the turning
points of the potential. It follows that $\alpha (x)$ generally oscillates
between the turning points. In practical implementations of amplitude-phase
formalisms, numerical methods aiming at minimizing the amplitude of the
oscillations have been devised. We have proposed in \cite{matz2000} such a
method based on the invariant and the nonlinear superposition principle in
the context of Ermakov systems. Here, the main point to be examined consists
in the relationship between non-oscillating amplitude-phase functions and
the functions obtained in the semiclassical $(\hbar \rightarrow 0)$ limit.\
More specifically, we will prove that in this limit, the only
non-oscillating solutions are the ones that yield classical quantities: in
particular, the only semiclassical phase function that does not oscillate is
the classical reduced action, and conversely the quantum continuation, for
finite $\hbar $, of the classical reduced action is a non-oscillating
function.

To this end, some properties of amplitude-phase functions, their behaviour
as a function of $x$ and $E,$ as well as the connection with Ermakov systems
will be recalled in Sec. 2. In Sec. 3, we shall discuss semiclassical
amplitude-phase functions by setting up a semiclassical Ermakov system; the
classical probability amplitude and reduced action will appear as a
particular solution of the semiclassical amplitude and phase functions.
Those results will then be employed in a formal $\hbar $ expansion of the
quantum amplitude-phase functions, to prove that provided the first order
functions are non-oscillating, the solutions to each order in $\hbar $ will
then not oscillate (Sec. 4). This will be followed in Sec. 5 by a discussion
of the results, in particular in relation to quantization of classically
integrable systems.

\section{Amplitude-phase functions}

Our concern here will be the 'auxiliary' amplitude and phase functions of
the Schr\"{o}dinger Eq. (\ref{3a}), with $p^{2}(x)=2m(E-V(x)),$ ie a
particle of mass $m$ and energy $E$ trapped in a potential well $V(x)$
having a single minimum. $V(x)$ is defined on an interval $]s_{1},s_{2}[$
(typically $s_{1,2}=\pm \infty $ or $0$). Atomic units will be used
throughout, except for the $\hbar $ factors which will be reestablished
where appropriate.

\subsection{Nonlinear superposition principle}

We collect in this paragraph the main results concerning amplitude-phase
functions that will be useful in what follows, omitting details (for more
details and the relevant references, see \cite{matz2000}). Eq. (\ref{3a})
defines a Sturm-Liouville problem, typically a singular problem on the half
line or real line when vanishing boundary conditions at $s_{1}$ and $s_{2}$
are implemented. However, our interest lies not in the specific
eigenfunctions or eigenvalues of the Schr\"{o}dinger equation, but in
relating linearly independent solutions of the linear equation to amplitude
and phase functions. We will denote by $u_{1}$ and $u_{2}$ two independent
solutions of Eq. (\ref{3a}) respectively regular at $s_{1}$ and $s_{2}$ and
with Wronskian $W=\mathcal{W[}u_{1},u_{2}]\equiv (\partial
_{x}u_{1})u_{2}-u_{1}(\partial _{x}u_{2})$ .

A general solution $u(x)$ of Eq. (\ref{3a}), which is readily written in
terms of the independent solutions $u_{1}$ and $u_{2}$, may also be obtained
as 
\begin{equation}
u(x)=b_{1}\alpha (x)\sin \left[ \phi (x)+b_{2}\right] ,
\end{equation}
where $b_{1}$ and $b_{2}$ are complex constants. A straightforward
substitution in Eq. (\ref{3a}) leads to the two equations 
\begin{eqnarray}
&&\partial _{x}^{2}\alpha (x)+p^{2}(x)\alpha (x)=\alpha (x)\left[ \partial
_{x}\phi (x)\right] ^{2}  \label{8a} \\
&&\alpha ^{2}(x)=\frac{a}{\partial _{x}\phi }  \label{8b}
\end{eqnarray}
where we can set $a^{2}=1$ without any loss of generality (since $a^{2}$ can
be absorbed into $\alpha $ by redefining $\alpha \rightarrow \alpha
/a^{1/2}),$ thus recovering Eq. (\ref{3b}). For obvious reasons, $\alpha $
and $\phi $ are known respectively as the amplitude and phase functions. In
terms of $u_{1}$ and $u_{2}$, it follows from standard results on Ermakov
systems that the general solution for $\alpha $ is given by 
\begin{equation}
\alpha (x)=\left[ \left( \frac{1}{2I}+2Ic^{2}\right) u_{1}^{2}(x)+\frac{2I}{%
W^{2}}u_{2}^{2}(x)-\frac{4Ic}{W}u_{1}(x)u_{2}(x)\right] ^{1/2},  \label{10}
\end{equation}
and the equation for the phase is readily integrated to give 
\begin{equation}
\phi (x)=\arctan \left[ \left( \frac{1}{2I}+2Ic^{2}\right) W\frac{u_{1}(x)}{%
u_{2}(x)}-2Ic\right] +\arctan 2Ic,  \label{11}
\end{equation}
where the integration constant is chosen so that $\phi (s_{1})=0$. Eq. (\ref
{10}) is an illustration of the nonlinear superposition principle \cite{ray
reid80}. $I$ and $c$ are two constants, independent of $x$ ($I$ is the
Ermakov, or Lewis-Ray-Reid, invariant). Note that the value of the phase
function at $s_{2}$, known as the accumulated phase, does not depend on the
constants $I$ and $c$ for the eigenvalues $E_{0}$ of the Sturm-Liouville
problem, since 
\begin{eqnarray}
\phi (s_{2},E &=&E_{0})=\pi n,  \label{11a} \\
\phi (s_{2},E &\neq &E_{0})=\arctan \left[ 2Ic(E)\right] +\frac{2n+1}{2}\pi ;
\label{11b}
\end{eqnarray}
here $n$ is an integer giving the number of nodes of $u_{2}.$ Note also that 
$\alpha (x_{1})$ is independent of $c$ if $x_{1}$ is a zero of $u_{1}$.

\subsection{Boundary conditions and energy dependence}

The boundary conditions for $\alpha $ and $\phi $ are therefore incorporated
through the parameters $I(E)$ and $c(E),$ which as indicated depend on the
energy. Normalization of the eigenfunctions $f$ of the Sturm-Liouville
problem with vanishing boundary conditions at $s_{1}$ and $s_{2}$, yields 
\begin{equation}
\int_{s_{1}}^{s_{2}}f^{2}(x)dx=I\partial _{E}\phi (s_{2},E=E_{0}).
\label{15c}
\end{equation}
By choosing the eigenfunctions to be normalized per unit energy increment, $%
I $ becomes an energy-independent positive constant, and amplitude and phase
functions depend on the single parameter $c(E)$.

We now introduce another solution $g(x)$ of the Schr\"{o}dinger Eq. (\ref{3a}%
), defined in terms of the solutions regular at $s_{1}$ and $s_{2}$ by 
\begin{equation}
g(x,c)=2I\left( \frac{u_{2}(x)}{W}-cu_{1}(x)\right) ,  \label{16}
\end{equation}
which fulfills $\mathcal{W}[u_{1},g]=2I$ and gives $\alpha (x)=\left[ \frac{1%
}{2I}\left( u_{1}^{2}(x)+g^{2}(x)\right) \right] ^{1/2}.$ This simply means
that $u_{1}$ and $g$ lag $\pi /2$ out of phase, and that with the
conventions of Sec. 2.1 ($a=1$ and $\phi (s_{1})=0),$ we have 
\begin{eqnarray}
u_{1}(x) &=&\sqrt{2I}\alpha (x,c)\sin \phi (x,c),  \label{17a} \\
g(x,c) &=&\sqrt{2I}\alpha (x,c)\cos \phi (x,c).  \label{17b}
\end{eqnarray}
We have emphasized the $c$-dependence of the different functions (though
self-consistency requires it, it may be checked explicitly that $u_{1}$ does
not depend on $c$).

\subsection{Oscillatory properties}

\subsubsection{Oscillations of the amplitude}

Let $t_{1}(E)$ ($t_{2}(E)$) be the inner (outer) turning point. For values
beyond the turning points (when $x<t_{1}$ or $x>t_{2}),$ we recast Eq. (\ref
{8a}) as 
\begin{equation}
\frac{1}{2}\left\langle \phi ;x\right\rangle =p^{2}(x)-\alpha ^{-4}(x),
\end{equation}
where $\left\langle \phi ;x\right\rangle \equiv \partial _{x}^{3}\phi
/\partial _{x}\phi -\frac{3}{2}(\partial _{x}^{2}\phi /\partial _{x}\phi
)^{2}$ denotes the Schwartzian derivative. Since $\alpha $ is a positively
defined quadratic form and $p^{2}(x)<0$ beyond the turning points, the
Schwartzian derivative of the phase is negative, and as it can be verified,
if $\left\langle \phi ;x\right\rangle <0$ on an interval then $\partial
_{x}\phi $ cannot have a positive local minimum on this interval, ie $\alpha 
$ cannot have a local maximum. By noting that $\alpha (x)\rightarrow +\infty 
$ when $x\rightarrow s_{1}$ and $x\rightarrow s_{2},$ we conclude that if $%
\alpha $ does not oscillate between the turning points, it will not
oscillate on the whole interval $]s_{1},s_{2}[$. Note that there is then a
unique value $x_{0},$ with $t_{1}<x_{0}<t_{2},$ such that $\partial
_{x}\alpha (x_{0})=0.$

However, as $u_{1}(x)$ and $u_{2}(x)$ oscillate for $t_{1}<x<t_{2}$, $\alpha 
$ will generally oscillate, by virtue of the nonlinear superposition
principle Eq. (\ref{10}), between the turning points, the local wavelength
being half that of $u_{1}$ or $u_{2}$. Nonetheless, at a specified energy,
there may be infinitely many values of $c$ giving a non-oscillating
amplitude function\footnote{%
For example, taking the second derivative of $\alpha ,$ it is seen that any
value of $c$ enclosed between $c_{\pm }(x)$, with $c_{\pm
}(x)=u_{2}(x)\left[ Wu_{1}(x)\right] ^{-1}\pm \left[ 2Iu_{1}(x)\right]
^{-1}\left[ 2I/p(x)-u_{1}^{2}(x)\right] ^{1/2}$ and where $x$ spans the
interval between the turning points, will do.}. We give in the following
paragraphs a sufficient (but not necessary) condition for $c(E),$ that is
the value of $c$ and its explicit energy-dependence, because this value has
remarkable properties related to classical quantities in the semiclassical
limit, as will be seen in Sec. 3.

\subsubsection{Inverted phase accumulation}

Proceeding as in Sect.\ 2.1, we can define a phase function $\bar{\phi}(x)$
such that $\bar{\phi}(s_{2})=0,$ ie the phase starts accumulating at $s_{2}$
instead of $s_{1}$. As can be easily seen, this amounts to exchange the
roles of $u_{1}$ and $u_{2}$; for instance, we now have $u_{2}=b_{2}\bar{%
\alpha}\sin \bar{\phi}$, where $\partial _{x}\bar{\phi}=\bar{\alpha}(x)^{-2}$
and $b_{2}$ is a constant to be set below. $\bar{\alpha}$ is given by 
\begin{equation}
\bar{\alpha}^{2}(x)=\left( \frac{1}{2I}+2I\bar{c}^{2}\right) u_{2}^{2}(x)+%
\frac{2I}{W^{2}}u_{1}^{2}(x)+\frac{4I\bar{c}}{W}u_{1}(x)u_{2}(x),
\end{equation}
with $\mathcal{W}[u_{2},u_{1}]=-W$ and where we assumed $I=\bar{I}$ for
simplicity. Generally $\alpha (x,c)$ and $\bar{\alpha}(x,\bar{c})$ are very
different functions (for example the $\bar{c}$ independent points of $\bar{%
\alpha}$ are now located at the zeros of $u_{2}$). Notwithstanding, it is
apparent that $\alpha (x,c)=\bar{\alpha}(x,\bar{c})$ iff $c^{2}=\bar{c}%
^{2}=W^{-2}-(2I)^{-2}$ and $\bar{c}=-c.$ We shall set 
\begin{equation}
c_{o}(E)=-\left[ \left[ W(E)\right] ^{-2}-\left[ 2I\right] ^{-2}\right]
^{1/2}.  \label{00}
\end{equation}
To keep the quadratic form real, this implies that $W^{2}<4I^{2}$, condition
to be assumed in the rest of the paper (this is not a problem in practice
because $\alpha $ and $\phi $ are left unchanged by the transformations $%
u_{1}\rightarrow \kappa u_{1},$ $W\rightarrow \kappa W,$ $I\rightarrow
\kappa ^{2}I,$ $c\rightarrow c/\kappa ^{2}$, so the Wronskian can be
conveniently rescaled).

Let us now suppose that $c=-\bar{c}=\pm c_{o}.$ We then have $\alpha =\bar{%
\alpha}.$ $b_{2}$ is found by evaluating $\mathcal{W}[u_{1},u_{2}]$ at $%
s_{2},$ which yields, by choosing a proper sign convention $b_{2}=\sqrt{2I}$%
. From Eqs. (\ref{16})-(\ref{17b}), it follows that 
\begin{equation}
\sin \bar{\phi}(\mp c_{0})=W\left[ \cos \phi (\pm c_{o})/2I\mp \left[
W^{-2}-(2I)^{-2}\right] ^{1/2}\sin \phi (\pm c_{o})\right] ,  \label{03}
\end{equation}
thereby obtaining the relation between $\phi $ and $\bar{\phi}$, which give
the oscillations of $u_{1}$ and $u_{2}$ only if $\alpha $ does not oscillate.

\subsubsection{Auxiliary quadratic form on the unit circle}

$\alpha ^{2}(x)$ is a positive definite quadratic form. Labelling $M$ the
matrix of the coefficients, we have $\det M=W^{-2}$ and $\mathrm{Tr\,}%
M=1/2I+2Ic^{2}+2I/W^{2}$. $\alpha ^{2}(x)$ can be reduced to the canonical
form 
\begin{equation}
\alpha ^{2}(x,c)=\lambda _{1}(c)v_{1}^{2}(x,c)+\lambda _{2}(c)v_{2}^{2}(x,c),
\label{04}
\end{equation}
where $\lambda _{i}(c)$ are the eigenvalues of $M$ ($\lambda _{1}\ge \lambda
_{2}$) and $v_{i}(x,c)$ are the eigenvectors, normalized so that $%
v_{1}^{2}(x,c)+v_{2}^{2}(x,c)=u_{1}^{2}(x)+u_{2}^{2}(x).$ We now introduce a
quadratic form $Q$ defined by 
\begin{equation}
Q(x,c)=\lambda _{1}(c)w_{1}^{2}(x,c)+\lambda _{2}(c)w_{2}^{2}(x,c),
\label{05}
\end{equation}
where $w_{i}^{2}(x,c)=v_{i}^{2}(x,c)\left[ u_{1}^{2}(x)+u_{2}^{2}(x)\right]
^{-1}$. As indicated, $Q,$ as well as the $\lambda _{i}$ and $v_{i}$ are $c$%
-dependent. $Q$ oscillates between its maximum and minimum values, which are
by construction given respectively by $\lambda _{1}$ and $\lambda _{2}.$

We now set $c=\pm c_{o}.$ Let $x_{1}$ ($x_{2})$ label the points where $u_{1}
$ ($u_{2}),$ vanishes; we then have $Q(x_{1,2},\pm c_{o})=2I/W^{2},$ so that
between 2 zeros of $u_{1}$ and $u_{2},$ $Q(x,\pm c_{o})$ has at least one
extremum on the unit circle (for those points, the equality $%
u_{1}^{2}(x)=u_{2}^{2}(x)$ is fulfilled). Note that $\alpha (x,c=\pm c_{o})$
goes through both the $c$-independent points of $\alpha (x,c)$ and $\bar{%
\alpha}(x,\bar{c}),$ so if $\alpha $ oscillates, then there is an extremum
of $\alpha $ between $x_{1}$ and $x_{2}$, and the sign of $\partial
_{x}\alpha $ alternates between the consecutive zeros of $u_{1}$ and $u_{2}$%
. This is illustrated on Fig. 1 for the specific case of the harmonic
oscillator (to be discussed in details in Sec. 5.4); the zeros of $u_{1}$
and $u_{2}$ are respectively shown as triangles and rectangles. Note also
that the maxima of $Q(c_{o})$ correspond to the minima of $Q(-c_{o})$, since 
$Q(-c_{o})=\bar{Q}(c_{o}).$ Combining Eqs. (\ref{04}) and (\ref{05}) and
taking the derivative $\partial _{x}\alpha (x_{1,2})$ as a function of $%
\partial _{x}Q,$ $Q,$ $u_{1}$ and $u_{2},$ it can be seen indeed that for $%
c=-c_{o}$ the sign of $\partial _{x}\alpha $ at 2 consecutive zeros, $x_{1}$
and $x_{2}$, alternates. However, for $c=c_{o},$ the sign of $\partial
_{x}\alpha $ between 2 consecutive arbitrary zeros of $u_{1}$ and $u_{2}$
does not change, and thus $\alpha (c_{o})$ does not oscillate\footnote{%
We noted, however, that given the behaviour of the amplitude function at $%
s_{1}$ and $s_{2},$ there is necessarily a point between $t_{1}$ and $t_{2}$
where $\partial _{x}\alpha $ vanishes. The argument sketched here relies on
the signs of the basis functions $u_{1}$ and $u_{2}$ and their derivatives
when $V(x)$ is monotonous on a full cycle of oscillation of the basis
functions, and excludes the neighbourhood around the bottom of the
potential, where $\partial _{x}\alpha $ changes sign (but $\partial
_{x}\alpha $ does not vanish exactly at the bottom of the potential, as
would be the case in the WKB approximation).}.

\subsubsection{Oscillations of the accumulated phase}

We have explained in \cite{matz2000} why obtaining non-oscillatory functions
is important when amplitude-phase methods are employed in scattering theory.
The goal there is to extend energy-normalization, which for the
eigenfunctions of the Sturm-Liouville problem is given by Eq. (\ref{15c}),
to functions $f(x)$ which converge at $s_{2}$ but diverge at $s_{1}$ (the
phase-shifted or scattered wavefunctions). By combining the continuity
equation for the probability density and L'H\^{o}pital's rule, improper
energy normalization is defined by 
\begin{equation}
\int_{r}^{s_{2}}f^{2}(x,E)dx=I\partial _{E}c\left( \frac{1}{2I}+2Ic
^{2}(E)\right)^{-1},  \label{13}
\end{equation}
where $r$ is a cut-off radius (and as above, $I$ is assumed to be
energy-independent). This normalization is of course arbitrary, since it is
governed by $c(E)$, but it conditions the energy-dependence of the different
scattering parameters. In particular, the accumulated phase, which is
unambiguously defined (Eq. (\ref{11a})) for the eigenfunctions of the
Sturm-Liouville problem, crucially depends (Eq. (\ref{11b})) on the
normalization when $E$ is not an eigenvalue.

More precisely, let us assume the eigenvalues of the Sturm-Liouville problem
to be given by $E_{0}=\xi (n),$ where $n$ is the number of zeros of the
corresponding eigenfunction (thus, of $u_{2}$), and $\xi (n)$ is an a-priori
arbitrary, but monotonous function admitting a differentiable inverse, $%
n(E)=\xi ^{-1}(E)$. The functional relation between $E_{0}$ and integer
values of $n$ is thereby extended to any energy $E$ lying between two
eigenvalues$,$ ie $E=\xi (n),$ $n$ real. The energy-dependence for $c(E)$
can now be chosen so as to extend the normalization of the eigenfunctions to
non-integer values of $n$ by equating Eq. (\ref{13}) to $I\pi \partial
_{E}\xi ^{-1}(E)$ (cf Eqs. (\ref{11a}) and (\ref{15c})), yielding 
\begin{equation}
c(E)=-\frac{1}{2I}\cot \pi \xi ^{-1}(E).  \label{13b}
\end{equation}
Substituting in Eq. (\ref{11b}) gives the following expression for the
accumulated phase: 
\begin{equation}
\phi (s_{2},E\neq E_{0})=\pi \xi ^{-1}(E)\equiv \pi n(E).  \label{13c}
\end{equation}
Thus the accumulated phase does not oscillate as a function of the energy
(it is a simple straight line as a function of $n$) and the value of $c$
given in Eq. (\ref{13b}) is the only value compatible with energy
normalization leading to a non-oscillating accumulated phase function. We
shall mention in Sec. 5.3 below the relation the specific form (\ref{13c})
has with the canonical action variable in classical mechanics. Note finally,
that provided the basis functions $u_{1}$ and $u_{2}$ are redefined so that
their Wronskian is proportional to $2I\sin \pi n(E),$ Eq. (\ref{13b})
becomes a particular form of the more general Eq. (\ref{00}): with such a
choice, the amplitude $\alpha $ is a non-oscillating function of $x$ and the
accumulated phase $\phi (s_{2})$ is a non-oscillating function of $E$.

\section{Semi-classical Ermakov system}

\subsection{Asymptotic solutions to the linear equation}

The approximate solutions to the one dimensional Schr\"{o}dinger equation
when $\hbar \rightarrow 0$ are well-known from the asymptotic theory of
ordinary linear differential equations \cite{fedoriuk93}. It follows from
Sec. 2.3.1 that it is sufficient to consider the solutions between the
turning points (ie for real $p(x)$). Real solutions are of the form 
\begin{equation}
\widetilde{u}(x)=\frac{a_{1}}{\sqrt{p(x)}}\sin \left[ \pm \int p(x^{\prime
})dx^{\prime }+a_{2}\right]  \label{12}
\end{equation}
where $a_{1}$ and $a_{2}$ are constants. Tilded ($\widetilde{\ })$
quantities will henceforth denote asymptotic (semiclassical) functions when
these are to be distinguished from the corresponding exact quantum
solutions. It is well-known from Hamilton-Jacobi theory that 
\begin{equation}
S(x,E)=\pm \int p(x^{\prime },E)dx^{\prime }+a_{2},  \label{12b}
\end{equation}
where $S(x,E)$ is known as the Hamilton-Jacobi characteristic function or
reduced action: the characteristics in phase-space are made up of the points 
$(x,\partial _{x}S)$.

\subsection{Ermakov system}

By direct substitution of a general asymptotic solution into the
Schr\"{o}dinger equation, and by labeling $\widetilde{\alpha }$ and $%
\widetilde{\phi }$ the semiclassical amplitude and phase functions, we
obtain a semiclassical Ermakov system 
\begin{eqnarray}
\hbar ^{2}\partial _{x}^{2}\widetilde{u}+\left[ p^{2}(x)+\frac{\hbar ^{2}}{2}%
\left\langle S;x\right\rangle \right] \widetilde{u} &=&0  \label{15a} \\
\hbar ^{2}\frac{\partial _{x}^{2}\widetilde{\alpha }}{\widetilde{\alpha }}%
+\left[ p^{2}(x)+\frac{\hbar ^{2}}{2}\left\langle S;x\right\rangle \right]
&=&\hbar ^{2}\left( \partial _{x}\widetilde{\phi }\right) ^{2},  \label{15b}
\end{eqnarray}
where again the bracket $\left\langle \;;x\right\rangle $ denotes a
Schwartzian derivative and we have, as for the usual amplitude-phase
functions $\widetilde{\alpha }^{2}=\widetilde{a}/\partial _{x}\widetilde{%
\phi }$ and thus $\partial _{x}^{2}\widetilde{\alpha }/\widetilde{\alpha }%
=-\left\langle \widetilde{\phi };x\right\rangle /2.$ Eqs. (\ref{15a})-(\ref
{15b}) are the semiclassical version of the quantum system given by Eqs. (%
\ref{3a})-(\ref{3b}). Eq. (\ref{15a}) is the modified Schr\"{o}dinger
equation exactly obeyed by the semiclassical wavefunctions, and Eq. (\ref
{15b}) is the nonlinear equation fulfilled by the semiclassical amplitude
function. The passage from the exact (quantum) Ermakov system to the
semiclassical one simply consists in a redefinition of the potential energy
function, and is identical if $\left\langle S;x\right\rangle $ vanishes.\
Although generally $\left\langle S;x\right\rangle $ is nonzero (except for
the free particle), $\left\langle S;x\right\rangle $ does tend to zero or to
a finite value in the limit of high quantum numbers (eg the harmonic
oscillator for the former, the centrifugal Coulomb potential for the
latter). Only if this value is negligible compared to the other terms in the
energy function does the high quantum numbers condition fit with the
semiclassical limit.

\subsection{General solutions}

$\widetilde{\alpha }$ and $\widetilde{\phi }$ are given in terms of two
independent functions $\widetilde{u}_{1}$ and $\widetilde{u}_{2}$ of Eq. (%
\ref{15a}) by the same relations, Eqs. (\ref{10}) and (\ref{11}), as in the
exact (quantum) case, with now tilded quantities. It is convenient, however,
to set the tilded constants $\widetilde{a},$ $\widetilde{I}$ and $\widetilde{%
W}$ equal to their quantum counterpart $a,$ $I$ and $W.$ This is done by
first noting that $\mathcal{W}[\alpha \sin \phi ,\alpha \cos \phi ]=a,$
which we then set equal to $\mathcal{W}[\widetilde{\alpha }\sin \widetilde{%
\phi },\widetilde{\alpha }\cos \widetilde{\phi }],$ so $\widetilde{a}=a=1.$
To preserve the Wronskians (cf Eqs. (\ref{17a})-(\ref{17b})), the
semiclassical function $\widetilde{u_{1}},$ of the form given by Eq. (\ref
{12}), is thus set as 
\begin{equation}
\widetilde{u_{1}}(x)=\sqrt{\frac{2I}{p(x)}}\sin S(x),  \label{19}
\end{equation}
where we have implicitly included $a_{2}$ in the reduced action so that $%
\widetilde{u_{1}}$ is the asymptotic approximation to $u_{1}$ in the
neighbourhood of an arbitrary $x$ lying between the turning points. An
independent solution $\widetilde{u_{2}}$ with Wronskian $\mathcal{W}[%
\widetilde{u_{1}},\widetilde{u_{2}}]=W$ is then obtained under the form $%
\cos (S(x)+b)/\sqrt{p(x)}$ as 
\begin{equation}
\widetilde{u_{2}}(x)=\sqrt{\frac{2I}{p(x)}}\kappa \cos \left[ S(x)+\arccos 
\frac{W}{2I\kappa }\right] ,  \label{20}
\end{equation}
where we have introduced the scaling factor $\kappa $ to keep all quantities
real. In what follows, we shall set $\kappa =1,$ which is tantamount to
rescaling $u_{2}$ (a similar rescaling was performed in the quantum case,
see below Eq. (\ref{00})).

\subsection{Non-oscillating solutions}

Substituting Eqs. (\ref{19})-(\ref{20}) in the expression for $\widetilde{%
\alpha }$ readily yields 
\begin{eqnarray}
\widetilde{\alpha }^{2} &=&\frac{1}{p(x)}\left[ 4I^{2}W^{-2}\sin
^{2}S(x)+\left( \cos S(x)-2Ic\sin S(x)\right) \right.  \nonumber \\
&&\times \left. \left\{ \cos S(x)-2I\left( c+W^{-1}\left[
4-W^{2}/I^{2}\right] ^{1/2}\right) \sin S(x)\right\} \right] ,
\end{eqnarray}
which is a highly oscillatory function for an arbitrary value of the
parameter $c$. However, it may be noted by inspection that for $c=-\left[
W^{-2}-(2I)^{-2}\right] ^{1/2},$ the oscillating terms are cancelled out.\
Remark that this is the same expression that was labelled $c_{o}$ in the
quantum case (Eq. (\ref{00})). Reestablishing $\hbar $, the amplitude now
reads $\widetilde{\alpha }^{2}(c_{o})=h/p(x),$ which given our assumptions
is a non-oscillating function of $x$. Identical substitutions may be done
for $\widetilde{\phi },$ from which it follows that the semiclassical phase
function is highly oscillatory for an arbitrary value of $c$ except if $%
c=c_{o}$, and in that case, 
\begin{equation}
\widetilde{\phi }(x,c_{o})=S(x)/\hbar .
\end{equation}

In short, non-oscillating functions are obtained for a unique value of the
parameter $c$, for which the semiclassical quantities match their classical
counterpart ($\sqrt{\hbar /p(x)}$ and $S(x)/\hbar $ are respectively the
classical probability amplitude and phase functions). Writing Eq. (\ref{15b}%
) as 
\begin{equation}
\frac{\hbar ^{2}}{2}\left[ \left\langle \widetilde{\phi };x\right\rangle
-\left\langle S;x\right\rangle \right] =p^{2}(x)-\frac{\hbar ^{2}}{%
\widetilde{\alpha }^{4}},  \label{20z}
\end{equation}
this means that each side of the nonlinear equation of the semiclassical
Ermakov system vanishes independently. Note also that $p^{2}(x)+\hbar
^{2}\left\langle S;x\right\rangle /2=\hbar ^{2}\left\langle \tan \widetilde{%
\phi }(c);x\right\rangle /2$ (this is established by using the M\"{o}bius
invariance of the Schwartzian derivative and establishing a $c$-dependent
linear transformation relating $\tan \widetilde{\phi }(x,c)$ to $\tan S(x)$).

\section{$\hbar $ expansions}

The link between the solutions of the quantum and semiclassical Ermakov
systems is done by employing a formal $\hbar $ expansion of the amplitude
function. As in the previous section, we assume $I$ and $W$ to be identical
in both the quantum and the semiclassical case. It is then straightforward
to show that to each order in $\hbar $ (as well as to infinite order),
non-oscillatory functions are obtained.

\subsection{Series expansions}

The formal asymptotic solution to the Schr\"{o}dinger equation for small
values of the parameter $\hbar $ is usually done by transforming it to the
Riccati form and then obtaining a recurrence relation between complex
function of order $j$ and the functions of lower order \cite{fedoriuk93}.
Here we proceed slightly differently, because we want the relations between
the amplitude and the phase to be verified to each order. We look for a
generic solution of Eq. (\ref{3a}$)$ under the form $u(x)=a(x)\exp
if(x)/\hbar $ where $a(x)$ and $f(x)$ are real functions admitting the
series expansions 
\begin{equation}
a(x)=\sum_{j=0}^{\infty }a_{j}(x)\hbar ^{j},\qquad f(x)=\sum_{j=0}^{\infty
}f_{j}(x)\hbar ^{j}.
\end{equation}
Substitution into the Schr\"{o}dinger equation gives $a_{j}(x)=0,$ $%
f_{j}(x)=0$ for odd $j$ and the following recurrence relations for even $j,$ 
$j\ge 2$: 
\begin{eqnarray}
&&\partial _{x}f_{j}(x)=\frac{1}{2a_{0}\partial _{x}f_{0}}\left[ \partial
_{x}^{2}a_{j-2}-\sum_{m=2}^{j-2}a_{m}\sum_{n=0}^{j-m}\partial
_{x}f_{n}\partial _{x}f_{j-m-n}\right] , \\
&&a_{j}(x)=\frac{b_{j}}{\sqrt{\partial _{x}f_{0}}}-\sum_{n=0}^{j-2}\frac{%
\int \left( 2\partial _{y}a_{n}\partial _{y}f_{j-n}+a_{n}\partial
_{y}^{2}f_{j-n}\right) /\sqrt{\partial _{y}f_{0}}dy}{2\sqrt{\partial
_{x}f_{0}}},
\end{eqnarray}
with $f_{0}(x)=\pm S(x),$ $a_{0}(x)=\pm \partial _{x}f_{0}^{-1/2},$ and
where the constants $b_{j}\ $appearing in the solution to the homogeneous
equations for $a_{j}(x)$ are all set to 0 for $j\ge 2$. The Wronskian
relations are then preserved to each order in $\hbar $ , that is 
\begin{equation}
\mathcal{W}[a(x)\sin f(x),a(x)\cos f(x)]=a_{0}^{2}(x)\partial
_{x}f_{0}(x)=\partial _{x}S(x)/p(x)=1.  \label{26}
\end{equation}

Between the turning points, the $\pm $ branches are combined to yield real
oscillatory functions. The formal expansion for $u_{1}$ (assuming again the
adequate integration constant to be included in $f_{0}$) is then given by 
\begin{equation}
u_{1}(x)=\sqrt{2I}\sum_{j=0}^{\infty }a_{j}(x)\hbar ^{j}\sin \left(
\sum_{i=0}^{\infty }f_{i}(x)\hbar ^{i}\right) .
\end{equation}
The expansion for the function lagging $\pi /2$ out of phase is trivially
obtained by using the $\cos $ function; from Eqs. (\ref{16}) to (\ref{17b})
the formal expansion for $u_{2}$ is then found as 
\begin{equation}
\fl u_{2}(x)=\frac{\sum_{j=0}^{\infty }a_{j}(x)\hbar ^{j}}{\sqrt{2I}}\left[
W\cos \left( \sum_{i=0}^{\infty }f_{i}(x)\hbar ^{i}\right) -I\left[
4-W^{2}/I^{2}\right] ^{1/2}\sin \left( \sum_{i=0}^{\infty }f_{i}(x)\hbar
^{i}\right) \right] .  \label{27}
\end{equation}
To first order these functions coincide by construction with the
semiclassical wavefunctions $\widetilde{u}_{1}$ and $\widetilde{u}_{2}.$

\subsection{Amplitude and phase expansions}

The $\hbar $ expansions for the amplitude and phase functions are obtained
by combining the nonlinear superposition principle [Eqs. (\ref{10})-(\ref{11}%
)] with the formal series expansions for $u_{1}$ and $u_{2}.$ The expansions
may be done to finite or infinite order. In the first case, the functions
are Taylor expanded around $\hbar =0$ after separating the classical terms $%
a_{0}$ and $f_{0}$. The infinite order case is analogous to the first order
case treated in Sec. 3.4. For example substituting the series expansions in
Eq. (\ref{11}) gives the expression (mod $\pi $) of the phase which can be
simplified as 
\begin{equation}
\fl \phi (x,c)=\mathrm{arccot}\left\{ \cot \left( \frac{S(x)}{\hbar }%
+\sum_{i=1}^{\infty }f_{i}(x)\hbar ^{i-1}\right) -\left[
2Ic(E)+IW^{-1}\left[ 4-W^{2}/I^{2}\right] ^{1/2}\right] \right\} .
\label{23}
\end{equation}
The amplitude function may be obtained by deriving this last equation,
keeping in mind the Wronskian relations (\ref{26}), yielding an expression
involving the sines and cosines of the expression between parentheses in Eq.
(\ref{23}). In both cases, the highly oscillatory terms are cancelled by
setting $c=c_{o}$ (in Eq. (\ref{23}) for example, the term between square
brackets then vanishes).

\section{Discussion}

\subsection{General Remarks}

We have thus seen that same value of $c$ gives rise to non-oscillating
functions both in the semiclassical and quantum cases. This is not
surprising, if the similarities between Eqs. (\ref{20}), (\ref{27}) and (\ref
{03}) on the one hand, together with the nonlinear superposition principle
in both the quantum and semiclassical Ermakov systems on the other hand are
considered: the same functional relation gives the amplitude-phase functions
in both cases. We note however that the simple arguments we have given
above, although physically appealing because of the direct connection to
standard classical quantities, are liable to a more rigorous treatment. For
example the series expansion obtained for the amplitude function does not
necessarily converge, and even its asymptotic properties as a function of
the parameter $c$ deserve a more thorough investigation. Actually, Lewis had
made the same remarks when studying the adiabatic invariant series of the
time-dependent classical harmonic oscillator in powers of an adiabatic
parameter $\varepsilon $ \cite{lewis68}, which is defined by the same
uncoupled Ermakov system as the one studied in this paper. Lewis's work was
the first (albeit implicit) application to Ermakov systems of Kruskal's
asymptotic theory of Hamiltonian systems extending the study of the
adiabatic invariants beyond the first order in $\varepsilon $ \cite
{kruskal62}. The transposition of these theories to the present problem is
not straightforward because we lack here the Hamiltonian formalism on which
these theories are based (eg, the integral invariants that appear in
Kruskal's theory would have here a rather obscure interpretation).
Nonetheless, the present results on the oscillatory properties can be
directly transposed to any uncoupled Ermakov system depending on an
'adiabatic' parameter $\varepsilon$ (which in the present context
corresponds of course to the Planck constant).

\subsection{Scattering basis functions}

Previously to the work of Fano \etal \cite{greene etal82}, the use of
amplitude-phase methods in scattering theory was limited to the high
kinetic-energy limit, and the derivatives of the phase $\phi $ above first
order neglected, thereby effectively restricting the treatment from the
start to the standard WKB approximation (see eg, Ch. 4.3 of \cite{rodberg
thaler67}). The more recent application of these methods to define a pair of
basis functions for phase-shifted wavefunctions in a potential (as outlined
above), relies on numerical treatments to minimize the oscillations; these
treatments are preferred even in the cases for which an analytic pair of
basis functions is known, such as the Whittaker functions for the
centrifugal Coulomb problem or the parabolic cylinder functions for the
harmonic oscillator. It is interesting to note that the approach suggested
in Sec. 2.3 yields, for an arbitrary potential with a single minimum, the
same relations that are known to be valid (and non-oscillating) for the
analytic functions in the mentioned special cases (eg in the Coulomb case
where the effective quantum number $\nu $ is defined by $\nu =\xi ^{-1}(E)+l$%
, the accumulated phase obtained with the Whittaker functions is $\phi
(\infty )=\pi (\nu -l),$ and improper normalization follows the
normalization of the eigenfunctions by normalizing to $\nu ^{3}/2,$
independently of $\nu $ being real or an integer \cite{seaton83}).

>From a formal standpoint, defining a specific basis of functions is
equivalent to defining the Green's function of the scattering process in the
asymptotic field. This is the Green's function that appears in the
Lippmann-Schwinger equations and through which the collision operators are
defined. This is why the collision operators depend on the parameter $c$.
Though at first sight this may appear as an unexpected feature, it must be
remembered that the explicit inclusion of closed channels leads to a
modification of the usual Green's functions through a term depending on the
accumulated phase \cite{matz99}.

\subsection{Semiclassical limit and classical quantities}

Standard semiclassical physics is usually not concerned by the quantum to
classical limiting procedure, because the strategy there is to start from
classical quantities at the outset and then proceed to quantization.
However, we have seen in Sec. 3 that the classical reduced action is only
one of the many phase functions that are obtained in the semiclassical limit
(namely the non-oscillating one). This type of problem is frequent in
'classicalization' procedures: a particular, often arbitrary choice has to
be made to recover classical quantities (action, Liouville equation, etc.) A
recent example is given in Ref. \cite{campos etal00}, where the passage from
Hilbert-space to classical-phase space operators involves particular choices
for the parameters in order to recover the classical dynamics. In other
works, this requirement takes the form of an additional ad-hoc condition
usually termed as the 'correspondence principle' (eg in Ref. \cite{leacock
padgett83} where a quantum version of the Hamilton-Jacobi equation is given
an additional boundary condition $p_{quantum}(x,E)\rightarrow
p_{classical}(x,E)$ when $\hbar \rightarrow 0$ and $E$ is fixed)\footnote{%
Other authors crudely suppress the $\hbar -$dependent terms in selected
equations where this suppression leads to classical relations (for example
this would be done in Eq. (\ref{3b}) by giving the amplitude squared the
dimensions of a classical quantity, getting thereby rid of $\partial
_{x}^{2}\alpha ,$ but would not be done in Eq. (\ref{3a}), which does not
support an obvious classical interpretation). This procedure has often been
criticized because the functions and the parameters appearing in the
equations depend on $\hbar $.}. Though much less general, our treatment is
more transparent in that the continuation of the classical reduced action in
the quantum domain is readily identified: it is the non-oscillating phase
function (and the continuation of the classical probability amplitude is the
non-oscillating amplitude function).

Other aspects of the classical-quantum correspondence for a classically
integrable and separable system deserve to be mentioned. The accumulated
phase (Eqs. (\ref{11a})-\ref{11b})) is seen to be directly related to the
line integral around a closed loop of $\alpha ^{-2}:$%
\begin{equation}
\oint \left[ \partial _{x}\phi (x,c)\right] dx=2\phi (s_{2},c).  \label{30}
\end{equation}
Contrary to EBK (torus) quantization, there are no caustics when dealing
with exact quantization, and the quantization condition reads $\oint \left[
\partial _{x}\phi (x,c)\right] dx=2\pi n,$ ie, $n$ is an integer and the
line integral does not depend on the particular value of $c$. Here we
understand by ''exact quantization'' the quantization of the exact quantum
phase, and not the exact WKB quantization of non-solvable potentials, as
employed by Voros \cite{voros00}. Note however that when $c=c_{o}$, the
(unquantized) integral (\ref{30}) reads, according to Eq. (\ref{13c}): 
\begin{equation}
\oint \left[ \partial _{x}\phi (x,c_{o})\right] dx=2\pi n(E).  \label{30b}
\end{equation}
Not only does $c=c_{o}$ preserve for non-integer real numbers the functional
relation valid for exact quantization, but it gives a parameterization of
the quantum equivalent of the canonical action variable.

In a similar vein, the period of motion $T$ is given in Hamilton-Jacobi
theory by taking the energy derivative of the reduced action along the
closed loop. The transposition to the phase function in the quantum case
would imply taking the energy derivative of Eq. (\ref{30}), which by Eqs. (%
\ref{11b}) and (\ref{13}) is proportional to the normalization. Again, the
normalization depends on $c(E)$ and to take $\partial _{E}\phi (c)$ as the
time parameterization doesn't appear to make much sense unless $c=c_{o},$
since any other value would lead to an oscillating function, which would
further not collapse to the classical period in the $\hbar \rightarrow 0$
limit.

\subsection{Example}

We illustrate the properties mentioned above on the harmonic oscillator, a
paradigm both in the Hamilton-Jacobi formulation of classical mechanics \cite
{goldstein80} and in the semiclassical theory of bound states \cite
{percival79}. The reduced action $S(x,E)$ of Eq. (\ref{12b}) is readily
obtained, from which it follows that 
\begin{eqnarray}
&&\partial _{x}S=\left( 2mE-m^{2}\omega ^{2}x^{2}\right) ^{1/2}  \label{40}
\\
&&J\equiv \oint \left[ \partial _{x}S\right] dx=2\pi E/\omega ,  \label{41}
\end{eqnarray}
where $m$ and $\omega $ are the mass and frequency of the oscillator; we
slightly depart from usual conventions and define $J$ to be the canonical
action variable. The period is recovered as $T=\partial _{E}J$. The standard
WKB solutions between the caustics, Eqs. (\ref{19})-(\ref{20}), are obtained
from these classical quantities. Semiclassical quantization must take into
account the singularities at the turning points, from which it follows that $%
E_{0}=\hbar \omega \left( n+1/2\right) ,$ with $n$ being an integer.

Quantum mechanically, the eigenvalues are given by $\xi (n)=\hbar \omega
\left( n+1/2\right) $ when $n$ is an integer that counts the zeros of the
eigenfunction, but this relation can be extended for any real value of $n$,
as discussed in Sec. 2.3.4. We then write $\xi ^{-1}$ as $n(E)=E/\hbar
\omega -1/2$. The derivative of the exact (quantum) phase is given by Eqs. (%
\ref{8b})-(\ref{10}) as 
\begin{equation}
\partial _{x}\phi (x,c)=\frac{m\omega }{\hbar }\left[ \left( \frac{1}{2I}%
+2Ic^{2}\right) u_{1}^{2}(x)+\frac{2I}{W^{2}}u_{2}^{2}(x)-\frac{4Ic}{W}%
u_{1}(x)u_{2}(x)\right] ^{-2}.  \label{42}
\end{equation}
Fig. 2 compares the classical conjugate momentum as given by Eq. (\ref{40})
with two exact phases $\hbar \partial _{x}\phi (x,c)$ obtained from Eq. (\ref
{42}) by using numerical solutions of the Schr\"{o}dinger equation $u_{1}$
and $u_{2}$ with respective vanishing boundary conditions at $x=-\infty $
and $x=+\infty $. One of the curves is for an arbitrary value of $c,$ the
other corresponds to $c=c_{o}$. It may be seen that even for a moderate
excitation ($n\approx 12$), the non-oscillating solution with $c=c_{o}$ can
barely be distinguished from the classical momentum at the same energy,
except near the turning points (the reason is that $\left\langle S;x\right\rangle$ is negligible, hence the exact quantum phase, solution of Eq. (\ref{8a}) tends to the semiclassical quantum phase $\widetilde{\phi }$ of Eq. (\ref{15b}), which is simply $S(x)$ when $c=c_{o}$).

The analog of the canonical action $J$ appears as the line integral (%
\ref{30}), which generally depends on $c$ \emph{except} when $n$ is an
integer in which case $\oint \left[ \hbar \partial _{x}\phi (x,c)\right]
dx=2\pi \hbar n.$ Thus for the quantized energies, the quantal line integral
differs from $J$ by the Maslov index. However, if 
$c=c_{o},$ we have for any $E$ [Eq. (\ref{30b})] 
\begin{equation}
\oint \left[ \hbar \partial _{x}\phi (x,c_{o})\right] dx=2\pi E/\omega
-\hbar /2,  \label{44}
\end{equation}
which is the classical result with an action correction coming from the
Maslov index. Note that taking the energy derivative of Eq. (\ref{30})
crucially depends, for \emph{any} energy $E$ (including the eigenvalues) on
the energy-dependence $c(E).$ Only the energy dependence given by the
relation (\ref{13b}) above, which yields Eq. (\ref{44}) in this case,
renders the usual relation for the period, and more generally follows the
classical time parameterization for conservative systems in Hamilton-Jacobi
theory. Thus, the parameter $c$, which appears free within quantum mechanics
and has no classical counterpart, must be constrained if the usual classical
relations for the oscillator are to be extended to quantum amplitude and
phase functions.

The usual interpretation of the invariant within Ermakov systems hinges on
the use of an original Hamiltonian or Lagrangian, from which the Ermakov
equations are derived. The invariant is then associated, by means of
Noether's theorem, with the conserved quantity of an auxiliary motion \cite
{simic00,ray and reid79}. Such an interpretation is of course not
available here, where the Ermakov equations are used in a quantum-mechanical
context. The invariant $I$ is employed in this context only to define the
normalization, through Eq. (\ref{15c}). However, for the choice $c=c_{o}$,
a further interpretative step may be taken, since then the term on the right
hand-side in Eqs. (\ref{15c}) or (\ref{13}) gives  
\begin{equation}
\frac{\hbar }{2m}I\partial _{E}\oint \left[ \hbar \partial _{x}\phi
(x,c_{o})\right] dx.
\end{equation}
For the harmonic oscillator, we have by Eq. (\ref{44}) and by adopting unity
normalization
\begin{equation}
I=\frac{m\omega }{\hbar \pi }.
\end{equation}
For other systems, $I$ usually depends on $n(E)$ except if the wavefunctions are energy-normalized. Elementary manipulations
yield the more general form
\begin{equation}
I=2\pi \left[ \int \lambda (x,E)dx\right] ^{-1},
\end{equation}
where $\lambda (x,E)$ is the local de Broglie wavelength and the range of
integration is restricted to the classical domain between the turning points.

\section{Conclusion}

Previous interest in amplitude-phase methods led us to investigate in this
work the oscillatory properties of the nonlinear equation of uncoupled
Ermakov systems. It was shown that non-oscillating amplitude-phase functions
in the space and energy variables have a particular feature in the
semiclassical limit: they yield classical quantities. We have seen that
although standard quantum-mechanical quantities, such as the wavefunctions
or the eigenvalues, are insensitive to the value of the parameter $c$ and
its energy dependence, there is a unique value of $c$ which appears as
connecting quantum amplitude and phase functions to their classical
counterpart. Only in this case can 'quantum characteristics' $(x,\partial
_{x}\phi )$ having a sense, and collapsing to $(x,\partial _{x}S)$ when $%
\hbar \rightarrow 0$ be defined. We insist again that from the point of view
of quantum mechanics, even in the semiclassical limit, this need not be the
case: any amplitude and phase functions obeying Eq. (\ref{20z}) will yield
correct semiclassical wavefunctions. A full study of these aspects on
specific physical systems will be given elsewhere. Further links with
current work on Ermakov systems may lead to a better appreciation, as well
as to an extension, of the formalism.

\ack
I thank T. Monteiro (UCL) for useful discussions on semiclassical physics.
Financial support from the European Commission's IHP-MCIF Programme is
acknowledged. 

\Figures

\begin{figure}
\caption{
$Q(c_{o})$ (broken line, left scale), $\alpha (c_{o})$ (non-oscillating
black line) and $\alpha (-c_{o})$ (oscillating grey line) [right scale] are
plotted for a harmonic oscillator (atomic units and $\omega =1$, $n(E)=4.4,$
see Sec. 5.4 for the definitions). All amplitude functions $\alpha (c)$ for
any $c$ go through the points $x_{1}$ (triangles), and any function $\bar{%
\alpha}(\bar{c})$ goes through the points $x_{2}$ (rectangles). Only the 2
functions $\alpha (\pm c_{o})$ go through both the points $x_{1}$ and $x_{2}$%
. Since $Q$ oscillates and $Q(x=x_{1,2},\pm c_{o})=2I/W^{2}$ is constant, $%
\partial _{x}Q$ has opposite signs at $x_{1}$ and $x_{2},$ as may be seen on
the figure. This is also the case for $\alpha (-c_{o})$ (grey line) which
therefore oscillates, but not for $\alpha (c_{o})$.}
\end{figure}

\begin{figure}
\caption{The positive branch of the classical canonical momentum $p(x)$ for a
harmonic oscillator ($\omega =1,$ atomic units) is plotted (grey line) vs
the ''quantal momentum'' $\partial _{x}\phi (x,c)$ \emph{i)} for an arbitray
value of $c$ (dashed line) and \emph{ii)} for the non-oscillating value $%
c=c_{o}$ (black line).}
\end{figure}

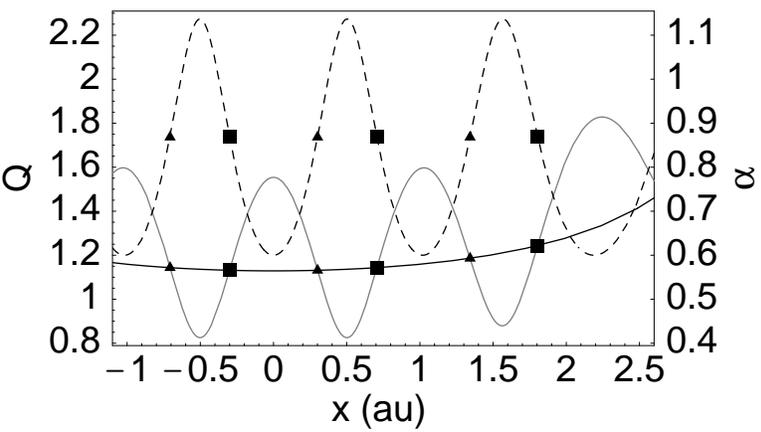
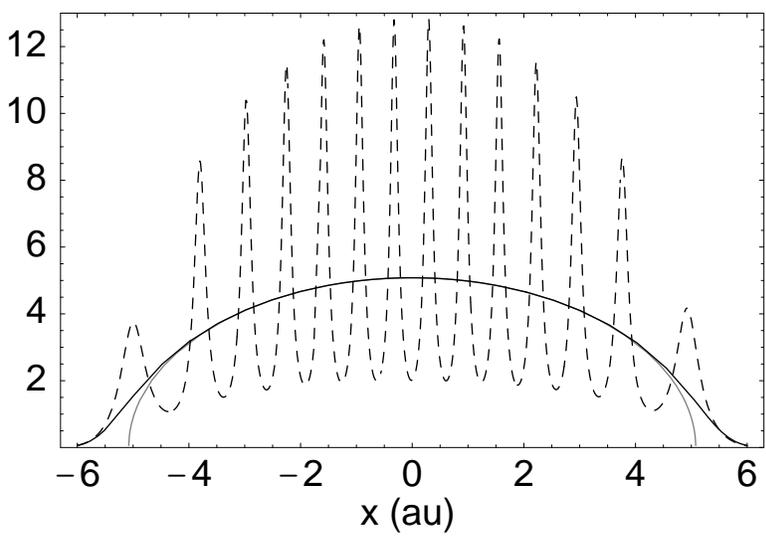
\end{document}